# 3D-printed conductive static mixers enable all-vanadium redox flow battery using slurry electrodes

Korcan Percin[a], Alexandra Rommerskirchen[a], Robert Sengpiel[b], Youri Gendel[c], Matthias Wessling[a,b]

[a]*DWI-Leibniz Institute for Interactive Materials*
*Forckenbeckstr. 50, 52074 Aachen, Germany*
[b]*RWTH Aachen University*
*Aachener Verfahrenstechnik-Chemical Process Engineering.*
*Forckenbeckstr. 51, 52074 Aachen, Germany.*
*Email: matthias.wessling@avt.rwth-aachen.de*
*Tel: +49 24180-95470; Fax: +49 24180-92252*
[c]*Technion-Israel Institute of Technology, Faculty of Civil and Environmental Engineering*
*Haifa 32000, Israel*

**Abstract**

State-of-the-art all-vanadium redox flow batteries employ porous carbonaceous materials as electrodes. The battery cells possess non-scalable fixed electrodes inserted into a cell stack. In contrast, a conductive particle network dispersed in the electrolyte, known as slurry electrode, may be beneficial for a scalable redox flow battery. In this work, slurry electrodes are successfully introduced to an all-vanadium redox flow battery. Activated carbon and graphite powder particles are dispersed up to 20 wt.% in the vanadium electrolyte and charge-discharge behavior is inspected via polarization studies. Graphite powder slurry is superior over activated carbon with a polarization behavior closer to the standard graphite felt electrodes. 3D-printed conductive static mixers introduced to the slurry channel im-prove the charge transfer via intensified slurry mixing and increased surface area. Consequently, a significant increase in the coulombic efficiency up to 80 % and energy efficiency up to 50 % is obtained. Our results show that slurry electrodes supported by conductive static mixers can be competitive to state-of-the-art electrodes yielding an additional degree of freedom in battery design. Research into carbon properties (particle size, internal surface area, pore size distribution) tailored to the electrolyte system and optimization of the mixer geometry may yield even better battery properties.

## 1. Introduction

All-vanadium redox flow batteries (VRB) have attracted considerable attention, since they are one of the most promising energy storage systems for large scale renewable energy applications [1–3]. Today, MWhs of energy storage capacity have been installed all over the world [4]. The power density of VRBs, however, is limited due to the slow kinetics of redox reactions on the electrode surface, thus shifting the focus towards investigating and improving electrode materials [5–8].

Electrodes for VRBs are commonly carbon-based materials [5]. Graphite felt (GF), for instance, is one of the most used electrode materials. GF benefits from a wide potential range, good stability, and low costs [9]. Numerous attempts were conducted to improve the performance of graphite electrodes such as surface modifications, usage of CNT electrodes, or the addition of noble metal catalyst layers [6–8]. A critical challenge remains the optimization of the active electrode surface area which affects the battery power output. A scalable electrode would provide freedom in designing the battery, thus an adjustable power output. Appleby et al. introduced the concept of slurry electrodes (flow electrodes) for zinc-air energy storage systems in 1976 [10].

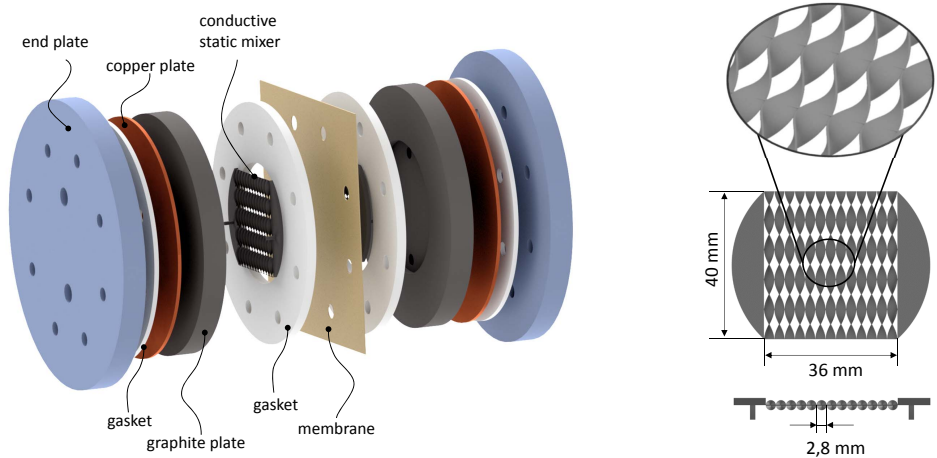

Figure 1: Design of the redox flow battery with (left) cell construction and (right) geometry of the static mixers.

Slurry electrodes are a dispersion of electrically conductive particles in a liquid electrolyte. These particles dynamically move inside the electrolyte and provide charge transfer from current collectors to the active species. Slurries were firstly mentioned as electrodes in 1928, cited by Heydecke and Beck [11]. Most early research about slurry electrodes focused on waste water treatment applications [12–14]. Later, slurry electrodes have also become interesting for energy storage applications, beginning with zinc-air batteries, lithium-ion batteries, capacitors and redox flow batteries (RFB) [10, 15–17]. Using slurry electrodes within a RFB electrolyte may be beneficial over conventional felt or foam materials, because they provide freedom of design. Slurry electrodes allow (1) scaling of the electrode surface area independent from the cross-sectional area of the cell, are (2) easy to produce and can be replaced without disassembling the cell and can (3) be easily recycled with filtration [18]. In this context, Hatzell et al. studied faradaic and capacitive effects in two different types of slurry electrodes on the positive electrolyte of the VRB, which has demonstrated a dramatic increase of energy and power density for $VO^2/VO_2^+$ redox reactions. Activated carbon (AC) and carbon nanotubes (CNT) were studied as slurry electrodes. AC was was identified as unsuitable due to the adsorptive surface properties. In contrast, CNT resulted in improved electrode characteristics [19]. However, CNTs may be expensive in such large-capacity batteries. A complete VRB using AC slurry electrodes was not accomplished yet.

Slurry electrodes show complex electronic properties correlated to rheological properties stemming from shear related particle-aggregation phenomena[20]. In this work, we hypothesize that conductive static mixers can positively affect the electronic and rheological properties of the slurry as they induce increased local shear rates. (Elucidating the fundamental correlations and the microscopic events goes beyond the scope of this contribution.) Such concept of 3D-printed static mixers and electrodes were introduced to electrochemical systems recently [21–24]. The static mixer reported here was 3D-printed, polymer-based with a conductive carbon layer deposited on its surface. The charge and discharge behaviour of the VRB was characterized via separate polarization studies. Furthermore, the battery was charged and discharged to analyze the cyclic performance. Results were compared to a conventional graphite felt electrode.

## 2. Materials and Methods

### 2.1. Cell Construction

In this work, a flow cell with a geometrical area of approx. 18 cm$^2$ was used, shown in Fig.1a. A fumapem$^®$ F-14000 (FuMA-Tech GmbH, Germany) served as the separating proton exchange membrane. The membrane was pretreated in 10 wt.-% HNO$_3$ solution for 3 h at 90$^o$C and subsequently in demineralized water for 1 h at 90$^o$C. Epoxy impregnated graphite plates (Müller & Rössner GmbH & Co., Germany) engraved with 2 mm channels distributed the electrolyte across the active area of the membrane and conducted the current to the copper current collectors. Graphite felt (GF) electrodes (4.6 mm thickness, GFD4.6EA, SGL Group, Germany) were used for both electrodes in a zero-gap configuration. The GF was pretreated at 400$^o$C for 16 h. To replace the GF, slurry electrodes were prepared by dispersing various amounts of activated carbon (Norit D10, D50=30 $\mu$m, 600 m$^2$ g$^{-1}$, Alfa Aesar, Germany) or graphite powder (synthetic, <20 $\mu$m, 20 m$^2$ g$^{-1}$, Sigma Aldrich, Germany) in the electrolyte.

Static mixers consisting of a single layer of twisted filaments with helical structure, adapted from Fritzmann et al. [25], were applied (Fig.1b). The static mixers were produced via additive manufacturing (Stratasys Objet Eden 260 V, RGD525) and pretreated with 75 % sulfuric acid for ten minutes to improve the coating adhesion. The static mixers were then spray-coated with a conductive graphite spray (Graphit, Cramolin). These conductive static mixers were used in combination with slurry electrodes to increase the contact area between particles and current collector and thus enabling a better charge transfer from the current collector surface to the slurry particles. Better designs with intensified slurry mixing may be applied and can be obtained from the heat exchange literature [26].

*2.2. Electrolyte Preparation*

A vanadium electrolyte solution (GfE Metalle und Materialien GmbH, Germany) was used as electrolyte for both half-cells. A peristaltic pump (Masterflex L/S, Cole Parmer) maintained a flow rate of 20 mL min$^{-1}$ for all experiments. For the preparation of catholyte and anolyte, volumes of 30 mL were initially charged and discharged with a constant current (CC) of 40 mA cm$^{-2}$. As a result, the vanadium ions were fully converted to $V^{3+}$ and $VO_2$ for the two half-cells, respectively. The slurry electrodes were dispersed in both half-cells after the initial charge and discharge. A steady nitrogen purge was used to prevent any oxidation of the vanadium ions.

*2.3. Electrochemical Measurements*

The measurements were performed with a potentiostat/galvanostat with impedance module (PGSTAT302N, Metrohm GmbH). Prior to each polarization experiment, an electrochemical impedance spectroscopy measurement was conducted to characterize the polarization curves by subtracting the ohmic resistances. For this purpose, high frequency resistances (HFR) were measured at $10-30\ kHz$ by using $10\ mV$ amplitude of alternating current potential. The ohmic resistance values were obtained from the Nyquist plot, where the spectrum intersects with the real axis. The area specific resistance (ASR) was calculated by multiplying the available membrane surface area with the HFR data. These values were used to correct the ohmic drop throughout the cell potential measurements.

The polarization curve experiments consisted of a series of galvanostatic charging steps, each time with a higher current density. CC was applied for $30\ s$ for every step, then the voltage data for the corresponding current density was recorded. The cutoff voltage was set to $1.9\ V$ for charging and $0.8\ V$ for discharging. The current density at these cutoff voltages is called cutoff current density in the following. At the end of each polarization curve, the battery was brought to the same open circuit potential (OCP) within a range of $\pm 30\ mV$ to maintain a similar state of charge. Following the polarization experiments, the battery was charged and discharged with a CC between $1.5\ V$ and $0.8\ V$ for ten cycles. All measurements were performed at room temperature with graphite felts or varied slurry particle concentrations, ranging from 0 to 20 wt.-%.

## 3. Results and Discussion

Polarization studies are performed to evaluate the cell resistances. Three main resistance regions can be recognized in these polarization studies. The first region is described as activation loss related, where the potential instantly increases from OCP to higher potentials. These type of losses are usually explained by electrode polarization and limited reaction kinetics. The second region represents ohmic resistance related losses where the potential shows linear behavior with increasing current. The main resistances in this region are electrolyte, membrane and electrode contact resistances. The third region is the transport region, where the voltage required to maintain the increased current changes to significantly higher values due to the diffusion-limited mass transfer of redox species towards and away from the electrode surface. At even higher current density, within the transport region, the battery approaches its maximum cutoff current density. High cutoff current densities are desired as they indicate low overpotentials and cell resistances. Ohmic resistance related losses can be estimated by HFR impedance analysis. With the ohmic resistance value subtracted from the polarization curve, the first and third polarization regions can be observed more accurately [27, 28].

The first experiment focuses on the effect of ohmic loss correction for the standard GF polarization. The ASR of the battery with GF electrodes is determined as approx. 2 $\Omega$ $cm^2$. This value is used to subtract ohmic

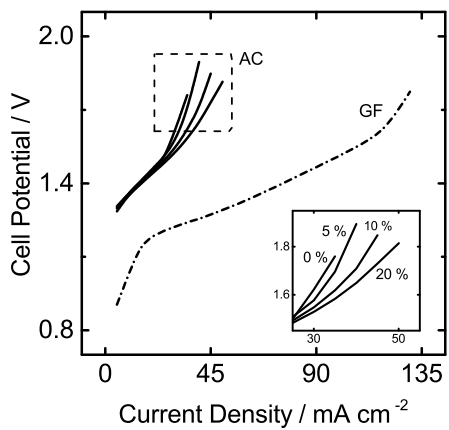

(a) Activated carbon 0, 5, 10, 20 % slurry

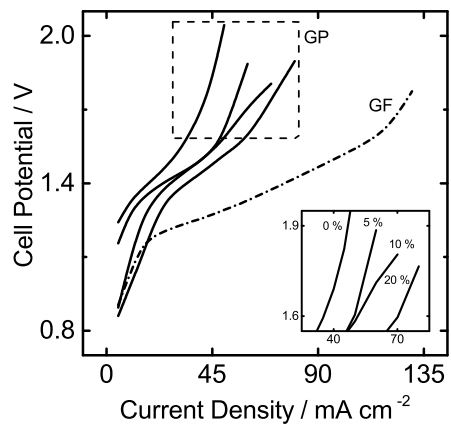

(b) Graphite powder 0, 5, 10, 20 % slurry

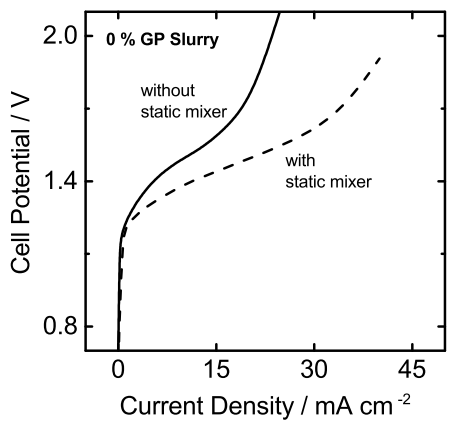

(c) 0 % graphite powder, effect of conductive static mixer

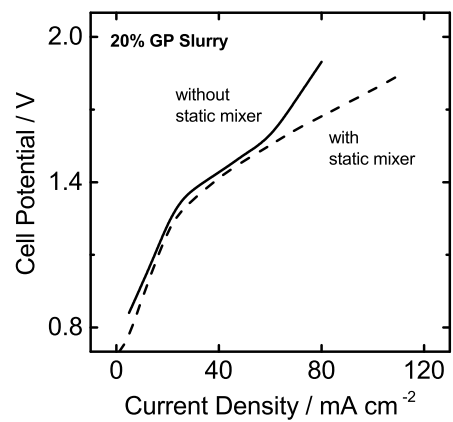

(d) 20 % graphite powder, effect of conductive static mixer

Figure 2: Charging polarization curves.

losses from the polarization curves. Ohmic correction shows a difference of $-0.2\ V$ at higher current densities. All the graphs included in this study are corrected for the ohmic resistance of each electrode setup and presented in form of cell potential.

Fig. 2a and 2b shows charging polarization curves of activated carbon (AC) and graphite powder (GP) slurries compared with the standard felt electrode. Three main polarization regions can be recognized for GF electrodes. The activation region expands up to around 15 $mA\ cm^{-2}$ and the third transport region becomes apparent after 120 $mA\ cm^{-2}$ until the cutoff current density of approx. 135 $mA\ cm^{-2}$. The achieved cutoff current density for the GF electrode is lower compared to previously published data [29] due to the non-optimized cell design. Yet, this study focuses on the question how slurry electrodes compare to a standard electrode material at a given cell design.

Fig. 2a presents the polarization behavior of AC slurry electrodes. The 20 % AC slurry shows the highest current density. Differences in AC content become apparent at high current densities. A high carbon content leads to more frequent particle contacts and thus improved charge transfer kinetics. However, the polarization curves show a significantly lower cutoff current density of AC slurries compared to the GF electrode. This phenomenon may be explained by the high porosity of the AC particles which adsorbs redox active species, thus resulting in lower current densities due to internal mass transport limitations [19]. Therefore, graphite powder is used as alternative to AC. GP particles provide higher conductivity, but less porosity and surface area compared to AC. Results are presented in Fig. 2b. The polarization curves of GP slurry electrodes also indicate a positive influence of increasing concentrations of particles. Furthermore, the curves are closer to the GF electrodes implying a better performance compared to AC slurries. Despite these promising results, such low current densities for the mass transport limitation region are not sufficient to run the battery with high-energy storage capacity. Therefore, conductive static-flow-mixers are introduced into the cell design. Static mixers disturb the particle flow and we hypothesize that they lead to more particle-current collector and particle-particle collisions.

Polarization studies using static mixers with a low number of conductive coating layers (through-plane resistance 1000 $\Omega\ cm$) do not show a significant difference compared to non-conductive static mixers. More layers of conductive coating material need to be applied until the through-plane resistance decreases to 100 $\Omega\ cm$. Such static mixers are applied in further polarization studies. Fig. 2c presents the charge polarization behavior of the flow battery operated with and without static mixers for particle-free electrolyte solutions. The polarization curve with static mixers indicates a better current/voltage performance even in a particle-free solution. At the same onset of about 1.25 V, current develops at a larger rate compared to the mixer-free solution. This may be attributed to the larger electrode surface of static mixers, which is not accounted for in the calculation of the current density. Starting from the region where mass transport limitations dominate, the application of static mixers leads to increased current densities due to an improved electrolyte flow distribution. Fig. 2d shows the polarization behavior of 20 % GP slurry electrodes with and without static mixers. Operating slurry electrodes in combination with static mixers leads to a reduction of mass transport limitations, while activation and ohmic region of both polarization curves are similar. The charge transfer along the battery is improved when particles collide with the current collector more often. Consequently, static mixers are used for further polarization experiments.

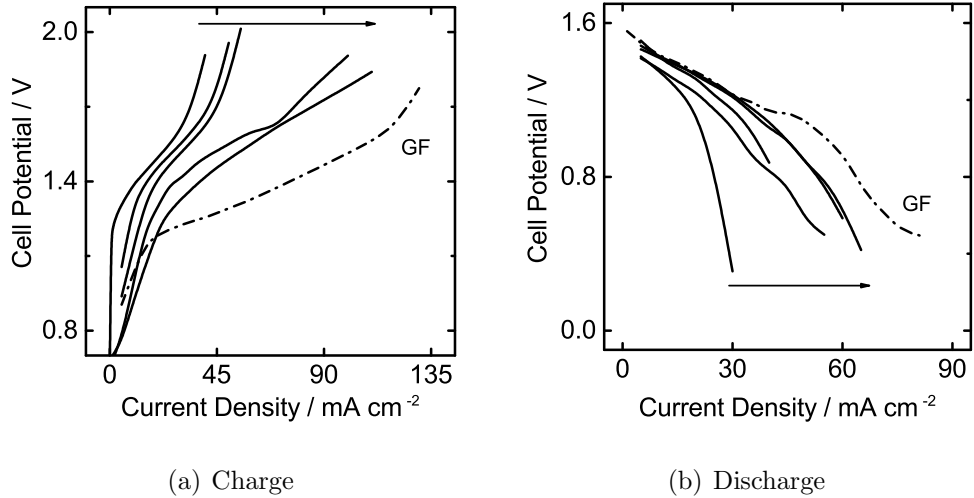

(a) Charge  (b) Discharge

Figure 3: Polarization curves with conductive static mixers, 0, 5, 10, 15 and 20 % slurry and graphite felt electrodes (Consecutive increase of slurry concentration indicated by direction of the arrows).

Fig. 3 demonstrates charge and discharge polarization of slurry electrodes with static mixers. GP slurry concentrations are 0, 5, 10, 15 and 20 %. Increasing slurry concentration leads to higher current densities and thus a performance closer to GF electrodes. Interestingly, Fig. 3a indicates a leap for 15 and 20 % slurry electrodes. A possible explanation for this may be the more frequent occurrence of percolations with increasing particle concentrations. Particles conduct more electrons from the current collector to the active species as result of percolating particles, which results in higher current densities. These higher numbers of particle percolations may be considered as transient networks of particles or transient agglomerates [30, 31]. A similar leap is not detected in Fig. 2b. Therefore the occurrence of agglomerations can be explained by the cell volume. Static mixers decrease the total available volume for the slurry flow resulting in more frequent agglomerations. Accordingly, a better charge transfer is achieved for highly concentrated slurries. Fig. 3b depicts the battery discharge behavior. Results obtained from discharge experiments indicate lower current densities for each slurry concentration and the GF. All discharge experiments illustrate less distinguishable polarization regions, which are dominated by the mass transport region. The 20 % slurry reaches almost 70 $mA\ cm^{-2}$ as discharge cutoff current density.

The experiments indicate that the 20 % slurry electrodes are most promising. However, the experiments were often interrupted due to clogging at tube fittings, caused by high viscosities and sedimentation of slurry particles. Therefore, further experiments are performed with 15 % slurry electrodes, which showed a similar behavior during the polarization studies (Fig. 2b). Following, charge-discharge cycles are performed with a current density of 20 $mA\ cm^{-2}$, which was chosen according to the second polarization region. The results are presented in Fig. S1. Minor capacity decay is visible from cycle to cycle. This can be explained by the crossover of vanadium ions through the membrane and may be avoided by choosing a better membrane [32–34]. Fig. 4 shows the coulombic (CE) and energy (EE) efficiency of ten charge-discharge cycles with and without static mixers. The CE is slightly above 50 % for the system without static mixers. Noticeably, the CE reaches up to 80 % when static mixers are implemented. A similar trend can be seen from the presented EE results as it reaches values above 50 % from originally 30 %. The voltage efficiency has values around 60 % with no significant difference for the operation with or without static mixers. The efficiencies are approximately constant for ten cycles. The increase of CE and EE can be explained by the improved transport of charges to the redox species due to static mixers.

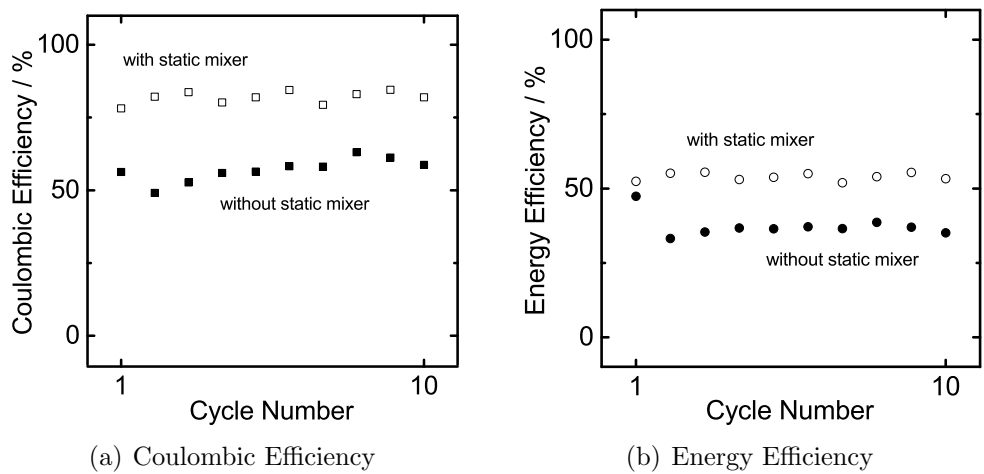

(a) Coulombic Efficiency  (b) Energy Efficiency

Figure 4: Cyclic charge-discharge efficiencies for 15 % GP at 20 $mA\ cm^{-2}$.

The increase in CE and EE correlates with the polarization study shown in Fig. 2d. A wider ohmic region range with belated mass transport limitation region results in higher efficiencies. This can be achieved by applying conductive static mixers, which leads to an increased current collector surface area and disturbed electrolyte-particle flow and thus improved charge transfer. Commercial VRBs with graphite felts usually achieve CEs higher than 90 % and EEs higher than 70 % [29]. Slurry electrodes used in this study reach lower efficiencies, which may be due to adsorptive properties of the particles. Adsorption of redox active vanadium in the pores of the particles leads to mass transfer limited polarization behavior and further side reactions. Our approach shows the possibility to use slurry electrodes in a VRB and the necessity to improve mixing and electrode/slurry contact with conductive static mixers. It is of utmost importance to further understand the effects of particle morphology and properties on the charge transfer and to improve the static mixer design.

## 4. Conclusions

In this study, a vanadium redox flow battery is operated with slurry electrodes. Activated carbon and graphite powder slurry electrodes are studied at particle concentrations up to 20 wt.-%. AC slurries do not show promising results whereas GP slurries exhibit similarities to the standard graphite felt electrodes. To improve the applicability of slurry electrodes, conductive static mixers are inserted into the electrolyte channels. Such system with conductive static mixers exhibits reduced mass transport limitations and an increased cutoff current density. Cyclic charge-discharge experiments indicate 80 % CE and 50 % EE when using static mixers.

**Acknowledgements**

This project has received funding from the European Research Council (ERC) under the European Unions Horizon 2020 research and innovation program (694946) and German Federal Ministry of Education and Research (BMBF) under the project $Tubulair\pm$ (03SF0436B). The authors would like to thank Kristina Heesen, Jan Keil, Dirk Pietzonka and Hansen Mou for their constructive effort to this study.

## Supplementary Material

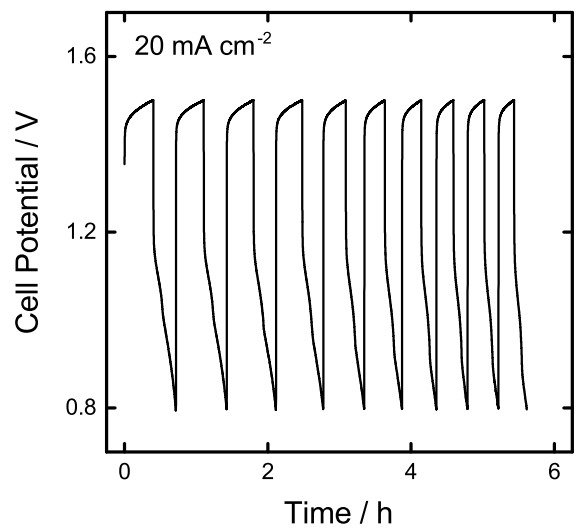

Figure S1: Charge-discharge cycles for 15 % GP.

## Graphical Abstract

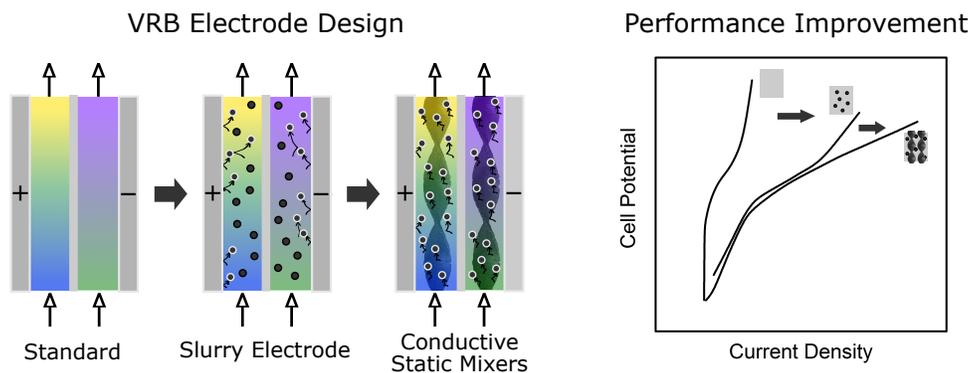

19